\documentclass[aps,prb,groupaddress,twocolumn,showpacs,amsmath,amssymb,amsfonts,floatfix]{revtex4-1}
\usepackage{graphicx,color,dcolumn,bm}
\usepackage{CJK}
\usepackage[colorlinks=true,citecolor=blue,urlcolor=blue]{hyperref}
\usepackage{natbib}

\begin{document}
\begin{CJK}{UTF8}{bsmi}
\title{Parity quantum numbers in the Density Matrix Renormalization Group}
\author{Yu-Chin Tzeng (曾郁欽)}
\email{d00222009@ntu.edu.tw}
\affiliation{Department of Physics, National Taiwan University, Taipei 106, Taiwan}
\affiliation{Center of Theoretical Sciences, National Taiwan University, Taipei 106, Taiwan}

\date{\today}
\begin{abstract}
In strongly correlated systems, numerical algorithms taking parity quantum numbers into account are used not only for accelerating computation by reducing the Hilbert space but also for particular manipulations such as the Level Spectroscopy (LS) method. By comparing energy difference between different parity quantum numbers, the LS method is a crucial technique used in identifying quantum critical points of Gaussian and Berezinsky-Kosterlitz-Thouless (BKT) type quantum phase transitions. These transitions that occur in many one-dimensional systems are usually difficult to study numerically. Although the LS method is an effective strategy to locate critical points, it has been lacked an algorithm that can manage large systems with parity quantum numbers. Here a new parity Density Matrix Renormalization Group (DMRG) algorithm is discussed. The LS method is the first time performed by DMRG in the S=2 XXZ spin chain with uniaxial anisotropy. Quantum critical points of BKT and Gaussian transitions can be located well. Thus, the LS method becomes a very powerful tool for BKT and Gaussian transitions. In addition, Oshikawa's conjecture in 1992 on the presence of an intermediate phase in the present model is the first time supported by DMRG.
\end{abstract}

\pacs{
02.70.-c,  
75.10.Pq,  
05.10.Cc}
\maketitle
\end{CJK}

\section{Introduction}
In one-dimensional (1D) quantum many-body systems, Berezinsky-Kosterlitz-Thouless (BKT)\cite{BKT} and
Gaussian type quantum phase transitions are usually difficult to be studied numerically.
New methods for accurate determination on the BKT and Gaussian critical points
have being proposed.\cite{BKTmethod,Zhou,MFYang,Tzeng,Hu2011,Rachel2012}
However, for examples, precise detections from entanglement entropy need to compute on very large
size $N>10000$ systems\cite{Hu2011}, and detections from bipartite fluctuations need the preliminary
knowledge of Luttinger parameters\cite{Rachel2012}.
The Level Spectroscopy (LS) method\cite{LS1997,LS1998} based on the sine-Gordon theory is an old
method for the BKT and Gaussian type quantum phase transitions.
By comparing different excitation energies, the LS method
is able to detect the Gaussian and BKT critical points accurately.
Ground state phase diagrams have been studied by the LS method in many models.%
\cite{Nakamura1999,Chen2003,Hijii2005,Furukawa2010,Tonegawa2011}
However, implementation of the LS method usually needs precise energy
eigenvalues with parity quantum numbers $p=\pm 1$.
Although parity or space inversion is usually a good quantum numbers in condensed matter physics,
the Exact Diagonalzation (ED)\cite{HQLin1990} was the only
accurate numerical algorithm with parity quantum numbers.
Therefore finite size effect usually interferes the LS method because of the limited sizes in ED.
In order to avoid the finite size effect in some difficult cases\cite{Tonegawa2011,Guo2011},
a precise algorithm with large sizes and parity quantum numbers is now desired urgently.\cite{SPT}

\begin{figure}[b]
\includegraphics[width=2.8in]{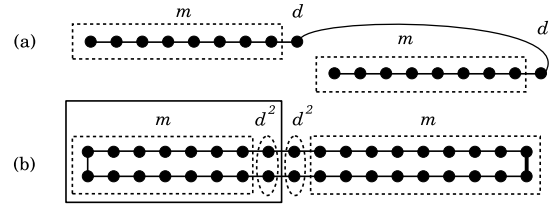}
\caption{(a) Parity DMRG scheme proposed by S\o{}rensen\cite{Sorensen1998}.
The left and right blocks must be the same size.
Parity quantum numbers are only available in the \textit{infinite-system DMRG} within this scheme.
$m$ is the dimension of a block and $d$ is the dimension of a site.
(b) Ladder scheme for parity quantum numbers in the sweeping procedure.
The left dash rectangle represents the \textit{system-block} and
the solid rectangle represents the \textit{enlarged-system-block}.
The dimension of one site (dash ellipse) becomes $d^2$.
The bold bond controls the boundary conditions.
Thus accurate results within open (OBC), periodic (PBC)
and twisted boundary conditions (TBC) can be easily obtained.
}\label{fig:scheme}
\end{figure}
Density Matrix Renormalization Group (DMRG)\cite{DMRG,White1996,White2005,RMP2005} is a powerful
numerical method of strongly correlated systems in 1D and two-dimensional (2D) lattices.
Lattice size in DMRG usually achieves hundreds of sites with high accuracy.
Recent study on spin-1/2 system in Kagome lattice revealed that DMRG may be the most accurate
numerical method in frustrated 2D systems.\cite{Yan2011}
Symmetry also plays an important role in DMRG.
U(1) symmetry, such as conservation of number of particles and
$z$-component of total spin, is the most frequently employed.
Non-Abelian symmetry\cite{McCulloch} and other descrete symmetries\cite{Sorensen1998,Ramasesha1996}
are often more complicated and rather difficult to implement.
For the parity symmetry, DMRG faces intrinsic difficulties in the algorithm.
The core problem of parity in DMRG is that the algorithm divides superblock into left and right parts.
This division destroys the spacial inversion if the length of left and right blocks are not equal.
When they are equal,
for example the \textit{infinite-system DMRG}, parity can be utilized.\cite{Sorensen1998}
However, \textit{finite-system DMRG} or so-called sweeping procedure
is usually needed for pushing ground state into a reliable precision.
In this case parity is difficult to be utilized.

In this paper, parity quantum numbers $p=\pm 1$ in the sweeping procedure
are discussed in an alternative ladder scheme.
The trick is very simple and no prior knowledge of group theory required.
Therefore the LS method overcomes the limitation of small sizes in ED
and becomes a very powerful tool in detecting BKT and Gaussian transitions.
In the state-of-the-art of DMRG, matrix-product state (MPS)\cite{Rommer1997}
is the main actor as variational algorithms.\cite{MPS}
However, the language of MPS is not used to describe the trick, but traditional DMRG instead,
since it continuously shows the significance in the studies.\cite{Yan2011,good}
In the following section, the trick is discussed in S=1/2 XXZ model.
In Sec.~\ref{sec:s2xxzd-a}, the proposed parity DMRG is applied to perform the LS method
in S=2 XXZ model with uniaxial anisotropy.
BKT and Gaussian transitions are both located precisely.
Thus in Sec.~\ref{sec:s2xxzd-b}, the boundaries of Intermediate-$D$ phase are determined.
In other words, the presence of Intermediate-$D$ phase
which Oshikawa predicted 20 years ago\cite{Oshikawa1992} is now supported by DMRG.
The conclusions are summarized in Sec.~\ref{sec:con}.

\section{Parity DMRG}
Consider the S=1/2 XXZ model,
\begin{eqnarray}\label{eq:xxz}
H=\sum_{j=1}^N\frac{1}{2}\left(S^{+}_jS^{-}_{j+1}+S^{-}_jS^{+}_{j+1}\right)+\lambda S^{z}_jS^{z}_{j+1},
\end{eqnarray}
where $\lambda$ is the anisotropic parameter and $N$ is total number of spins.
In developing numerical methods of strongly correlated systems, this model can be regarded as
a basic testing model\cite{HQLin1990,DMRG} because the dimension of one site $d=2$
and it can be solved by using Bethe ansatz.\cite{CNYang1966,Alcaraz1987}
The model undergoes a BKT quantum phase transition at $\lambda_c=1$.
Ground state energy per site at this critical point in the thermodynamic limit is $e_0=-\ln 2+\frac{1}{4}$.
Parity quantum numbers of finite size ground state and first excited state depend on $N$.
When $L=N/2$ is even (odd), the parity of ground state is also even (odd) and the first excited state is odd (even).
These two states become the two-fold degenerate ground state in the thermodynamic limit.

The reason that quantum numbers of $S_{tot}^z$ can be utilized is the fact that
reduced density matrix of \textit{enlarged-system-block} $\rho_{sys}$
and the $z$-component of total spin of \textit{enlarged-system-block} $S_{sys}^z$
are compatible, \textit{i.e.}, $[\rho_{sys},S_{sys}^z]=0$.
Thus each eigen-state of reduced density matrix has a quantum number of $S_{sys}^z$,
and these quantum numbers can be used for the next iteration.
However the parity operator $\mathcal{P}$ is a global operator
which only acting on the whole chain, the superblock.
It is impossible that in the original scheme to have an operator $\mathcal{P}_{sys}$
which only acting on the \textit{enlarged-system-block}.

In order to make it possible, the ladder scheme should be employed.
As shown in Fig.~\ref{fig:scheme}(b), the chain is folded as a two-leg ladder.
Now the global parity operator $\mathcal{P}$ can be regarded as an operation of exchanging legs.
Moreover,
\begin{equation}\label{eq:pi}
\mathcal{P}=\prod_{i=1}^{L}\mathcal{P}_i
\end{equation}
where $L=N/2$ and each $\mathcal{P}_i$ only exchanges legs of rung $i$.
Thus $\mathcal{P}_{sys}$ is the parity operator which only exchanges legs of
the \textit{enlarged-system-block} and
$[\rho_{sys},\mathcal{P}_{sys}]=[S_{sys}^z,\mathcal{P}_{sys}]=[\rho_{sys},S_{sys}^z]=0$.
The eigen-states of reduced density matrix have both quantum numbers of $S_{sys}^z$
and $\mathcal{P}_{sys}$ which can be used for the next iteration in sweeping procedure.
The bases of a rung, dashed ellipse in Fig.~\ref{fig:scheme}(b),
$|\tau_i\rangle\in\lbrace
\mid\uparrow\uparrow\rangle,
\mid\uparrow\downarrow\rangle,
\mid\downarrow\uparrow\rangle,
\mid\downarrow\downarrow\rangle\rbrace$.
Each $|\tau_i\rangle$ is an eigen-state of $S_i^z$, however,
these bases are not eigen-states of the parity $\mathcal{P}_i$.
Since $[\mathcal{P}_i,S_i^z]=0$, one can easily find the basis set
$\lbrace |\sigma_i\rangle\rbrace$ which are mutual eigen-states of $\mathcal{P}_i$ and $S_i^z$.
\begin{equation}\label{eq:basis}
|\sigma_i\rangle\in\left\lbrace
\frac{\mid\uparrow\downarrow\rangle-\mid\downarrow\uparrow\rangle}{\sqrt{2}},
\frac{\mid\uparrow\downarrow\rangle+\mid\downarrow\uparrow\rangle}{\sqrt{2}},
\mid\uparrow\uparrow\rangle,\mid\downarrow\downarrow\rangle\right\rbrace.
\end{equation}
These states are so-called singlet and triplet states
with corresponding quantum numbers of parity $p_i=-1,1,1$, and $1$, respectively,
and quantum numbers of $z$-component of spin $s_i^z=0,0,1$, and $-1$, respectively.
Eq.~(\ref{eq:pi}) implies the quantum number of \textit{enlarged-system-block}
$p_k=p_l p_n$, where $p_l$ is quantum number of \textit{system-block} and
$p_n$ is quantum number of a single rung.
These indexes satisfy the relation $k=(l-1)d^2+n$.

\begin{table}[b]
\caption{\label{tab:ge}%
Energy per site of 1D spin-1/2 XXZ model.
$p_0$ and $p_1$ are parity quantum numbers
of the ground state and first excited state, respectively.
The number of state kept in the largest size is up to $m=1500$
and the truncation error is of the order $10^{-9}$.
Three sweeps are performed.
PBC is used here, \textit{i.e.},
$H_1=H_L=\mbox{diag}(\frac{-2-\lambda}{4},\frac{2-\lambda}{4},\frac{\lambda}{4},\frac{\lambda}{4})$
in Eq.~(\ref{eq:ladder}).}
\begin{ruledtabular}
\begin{tabular}{rcrcrcr}
$N$ & $\lambda$ & $p_0$ & $E_0/N$ & $p_1$ & $E_1/N$ \\
\colrule
$54$ &1   &$-$& $-0.44343001$ &$+$& $-0.44189435$ \\
$100$&1   &$+$& $-0.44322957$ &$-$& $-0.44277665$ \\
$162$&1   &$-$& $-0.44317856$ &$+$& $-0.44300474$ \\
\end{tabular}
\end{ruledtabular}
\end{table}

\begin{figure}[t]
\includegraphics[width=2.5in]{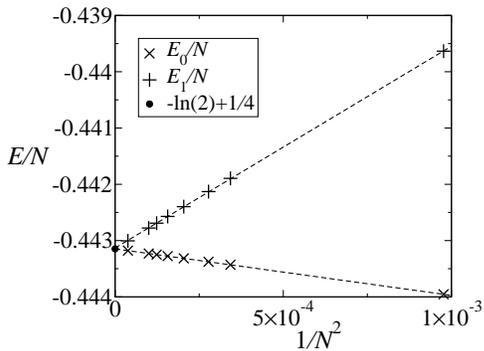}
\caption{The ground state and first excited state energy per site of spin-$\frac{1}{2}$ Heisenberg model,
\textit{i.e.}, $\lambda=1$ in Eq.~(\ref{eq:ladder}). PBC is used and three sweeps are performed in parity DMRG.
The dot indicates the exact ground state energy in the thermodynamic limit from Bethe ansatz.}%
\label{fig:eng}
\end{figure}

The $\lbrace|\sigma_i\rangle\rbrace$ forms a transformation matrix $Q$.
\begin{equation}
    Q=\left(\begin{array}{cccc}
    0 & 0 & 1 & 0 \\
    \frac{1}{\sqrt{2}} & \frac{1}{\sqrt{2}} & 0 & 0 \\
    \frac{-1}{\sqrt{2}} & \frac{1}{\sqrt{2}} & 0 & 0 \\
    0 & 0 & 0 & 1
    \end{array}\right).
\end{equation}
The spin operators of the first and second site of rung $i$ are
$S_{1,i}^z=Q^T(S^z\otimes 1_{d\times d})Q$ and
$S_{2,i}^z=Q^T(1_{d\times d}\otimes S^z)Q$,
and so do $S_{1,i}^+$, $S_{2,i}^+$, $S_{1,i}^-$ and $S_{2,i}^-$.
$1_{d\times d}$ is a $d\times d$ identity matrix.
The two spins of the first rung interact with each other in the ladder scheme.
The Hamiltonian of the first rung $H_1$ is a $d^2 \times d^2$ matrix and
it is fortunately already diagonalized with
eigenvalue $\frac{-2-\lambda}{4}$ for singlet state and
eigenvalues $\frac{2-\lambda}{4}$, $\frac{\lambda}{4}$ and $\frac{\lambda}{4}$
for triplet states, respectively.
The Hamiltonian of the latest rung $H_L$, the bold bond shown in Fig.~\ref{fig:scheme}(b),
can be flexible chosen as different boundary conditions.
For example,
$H_L=H_1=\mbox{diag}(\frac{-2-\lambda}{4},\frac{2-\lambda}{4},\frac{\lambda}{4},\frac{\lambda}{4})$
is the periodic boundary condition (PBC),
$H_L=0$ is the open boundary condition (OBC),
and $H_L=\mbox{diag}(\frac{2-\lambda}{4},\frac{-2-\lambda}{4},\frac{\lambda}{4},\frac{\lambda}{4})$
is the twisted boundary condition (TBC).
Thus Eq.~(\ref{eq:xxz}) can be rewritten as
\begin{eqnarray}\label{eq:ladder}
    H&=&\sum_{\alpha=1}^2\sum_{i=1}^{L-1}\frac{1}{2}(S_{\alpha,i}^+S_{\alpha,i+1}^-
    +S_{\alpha,i}^-S_{\alpha,i+1}^+)+\lambda S_{\alpha,i}^zS_{\alpha,i+1}^z \nonumber\\
    &+& H_1 + H_L,
\end{eqnarray}
where $L=N/2$.
Once a DMRG programmer has above information, the DMRG with parity quantum numbers can be performed.
The energies of ground state and first excited state with corresponding parity quantum numbers
are listed in Tab.~\ref{tab:ge} and plotted in Fig.~\ref{fig:eng}
for looking at convergence of energy with respect to system size.

Although parity reduces the dimension of superblock by a factor 2,
the disadvantage of the ladder scheme is the dimension of one site becomes squared.
In other words, 4 in S=1/2 chain, 9 in S=1 chain or $t$-$J$ model,
16 in Hubbard model and 25 in S=2 chain.
The overall computational difficulty increases, especially when $d$ is very large.
One can further use single center site\cite{White2005}
to reduce the dimension of superblock. As mentioned by White,
the ladder scheme is a better configuration for PBC
but it only improves convergence with the number of sweeps.\cite{White2005}
I emphasize the feasibility of parity DMRG in the ladder scheme.
Thus in the case $d^2=25$, the S=2 XXZ model with uniaxial anisotropy is examined
in Sec.~\ref{sec:s2xxzd} as practical examples.
DMRG with a single center site\cite{White2005} does not be employed here.
Besides quantum numbers ($S_{tot}^z$ and $\mathcal{P}$), the only optimization in this work is the
\textit{wave-function transformation}\cite{White1996}
which proposed by White in 1996 becomes a standard optimization in performing DMRG.

\section{S=2 XXZ model with uniaxial anisotropy}\label{sec:s2xxzd}
S=2 spin chains are current interesting research topics, including topological phases and
quantum phase transitions\cite{Hirano2008,Jiang2010,Zang2010,Zheng2011,Tonegawa2011,Tu2011,Pollmann2012},
magnetization process\cite{Lou2000}, and cold atoms loaded into a 1D optical lattice\cite{PCChen2012}.
The S=2 XXZ model with uniaxial anisotropy is defined by
\begin{equation}\label{eq:s2xxzd}
H=\sum_{j=1}^N(S^{x}_jS^{x}_{j+1}+S^{y}_jS^{y}_{j+1}+\lambda S^{z}_jS^{z}_{j+1})+D\sum_{j=1}^N(S_j^z)^2,
\end{equation}
where $\lambda$ and $D$ are the XXZ anisotropy parameter and uniaxial anisotropy parameter, respectively.
This model exhibits a Haldane gap in the Haldane phase with a non-local string order.\cite{Haldane,Qin2003}
Although the ground state phase diagram for the entire parameter space is unclear,
it is expected to have Haldane, N\'eel, Large-$D$, Intermediate-$D$, XY1, XY4, and Ferromagnetic phases.
Quantum phase transitions with various universality classes are in this model.
It is still a controversial issue\cite{Oshikawa1992,Schollwock1995,Schollwock1996,noID,Hirano2008,Tonegawa2011}
that whether the Intermediate-$D$ (ID) phase which was conjectured by Oshikawa\cite{Oshikawa1992}
twenty years ago exists in ground state phase diagram.
Although it was concluded absence of ID phase by DMRG\cite{Schollwock1995,Schollwock1996,noID},
surprisingly by using ED to perform LS on very small sizes $N\leq 12$,
Tonegawa \textit{et al.} obtained the ID phase in a very narrow region.
However, the ED estimations may not reveal results of thermodynamic limit
because of finite size effect, especially when $D>2$.
Energy level crossing does not appear in small sizes at all.
The region of the phase diagram that can not be determined by ED
was thus determined by extrapolation of phase boundaries.\cite{Tonegawa2011}

\subsection{Level Spectroscopy method}\label{sec:s2xxzd-a}
Therefore the first application of parity DMRG is to perform the LS method by focusing on
the BKT transition from XY1 to Large-$D$ phase as well as
the Gaussian transitions from Large-$D$ to ID phase and ID to Haldane phase.
Following the LS method, these three excitation energies
$E_0(N,0,+;\mbox{tbc})$, $E_0(N,0,-;\mbox{tbc})$, and $E_0(N,2,+;\mbox{pbc})$ should be compared.
Where $E_0(N,M,p;\mbox{tbc})$ and $E_0(N,M,p;\mbox{pbc})$ denote the lowest energy eigenvalues
of $N$ spins in the subspace of $z$-component of total spin $M$
and parity quantum number $p$ within TBC and PBC, respectively.

\begin{figure}[t]
\includegraphics[width=3.4in]{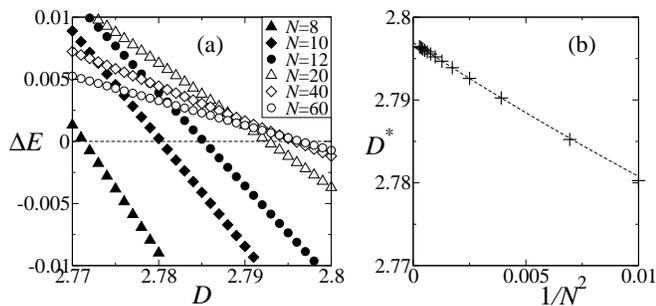}
\caption{
The LS method for BKT transition in the model Eq.~(\ref{eq:s2xxzd}).
$\lambda=0$ is fixed.
Energy differences in Eq.~(\ref{eq:bkt}) with different sizes are shown in (a)
and the extrapolation of critical point is shown in (b).
$N$=8, 10, and 12 are computed by ED, and larger sizes are computed by DMRG.
Only $N\geq 16$ are used for the extrapolation.
The XY1-Large-$D$ critical point from the extrapolation $D_c=2.796917\pm 5\times 10^{-6}$.
}\label{fig:BKT}
\end{figure}

The BKT pseudo-critical points of XY1-Large-$D$ satisfy the condition\cite{Tonegawa2011}
\begin{equation}\label{eq:bkt}
     E_0(N,0,+;\mbox{tbc})-E_0(N,2,+;\mbox{pbc})=0,
\end{equation}
and the Gaussian pseudo-critical points of Large-$D$-ID and ID-Haldane
satisfy the condition\cite{Tonegawa2011}
\begin{equation}\label{eq:id}
     E_0(N,0,+;\mbox{tbc})-E_0(N,0,-;\mbox{tbc})=0.
\end{equation}
These pseudo-critical points are extrapolated to $N\rightarrow\infty$
by assuming the quadratic scaling function of $N^{-2}$ after collecting the finite size results.
Fig.~\ref{fig:BKT} demonstrates the LS method for the BKT transitions.
$\lambda=0$ is fixed in Eq.~(\ref{eq:s2xxzd}).
The number of states kept is up to $m=800$, and truncation error is about $10^{-8}$.
BKT quantum phase transitions usually difficult to precisely locate numerically.
However, this critical point is well located by the LS method after using the parity DMRG.
As shown in Fig.~\ref{fig:BKT}(b), the critical point $D_c=2.796917\pm 5\times 10^{-6}$ is determined precisely.
By comparing the accuracy of recent development of detecting BKT quantum phase transitions,\cite{Rachel2012}
the LS method becomes a very powerful tool for the BKT quantum phase transitions.

\begin{figure}[t]
\includegraphics[width=2.6in]{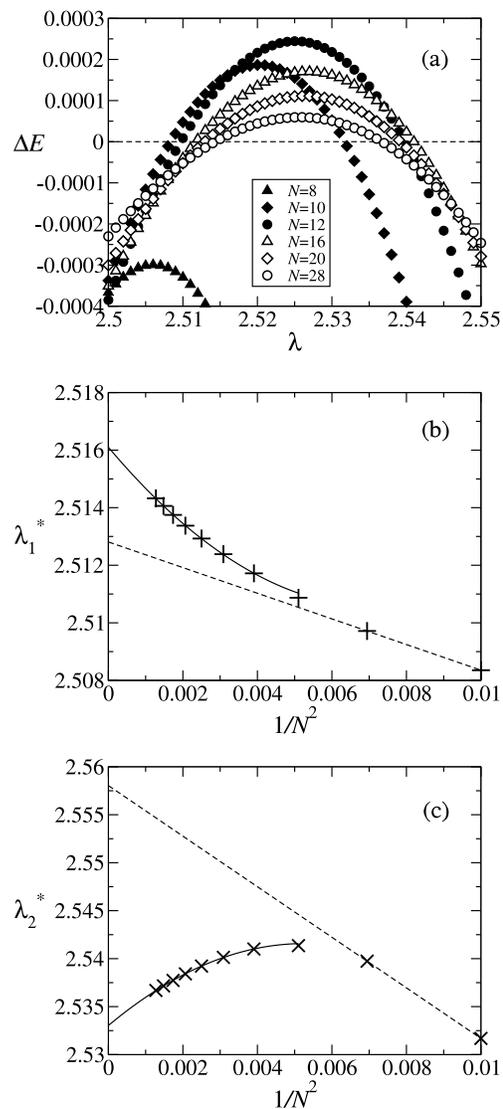}
\caption{The LS method for Gaussian transition in model Eq.~(\ref{eq:s2xxzd}).
$D=2.1$ is fixed.
Energy differences in Eq.~(\ref{eq:id}) with different sizes are shown in (a)
and the extrapolations of critical points are shown in (b) and (c).
$N\leq 12$ are computed by ED, and larger sizes are computed by DMRG.
Since energy level crossing takes place when $N\geq 10$,
it is not sufficient for quadratic fitting with only the ED data.
The dash lines in (b) and (c) are linear guild from ED.
The solid lines are the extrapolations with only $16\leq N\leq 28$.
The critical points for Large-$D$-ID and ID-Haldane are
$\lambda_{c1}=2.51610\pm 3\times 10^{-5}$ and $\lambda_{c2}=2.53303\pm 4\times 10^{-5}$, respectively.}\label{fig:LS}
\end{figure}
For the Gaussian transitions, Fig.~\ref{fig:LS} shows the LS results with fixed $D=2.1$.
Due to the finite size effect, the energy level crossing does not take place until $N\geq 10$.
Consequently, it is not sufficient for quadratic fitting with only the ED data of $N=10$ and $12$.\citep{Tonegawa2011}
In fact, the ID-Haldane pseudo-critical point $\lambda_2^\ast$ behaves non-monotonically with size.
As shown in Fig.~\ref{fig:LS}(b) and (c), only $16\leq N\leq 28$ are used for the extrapolation.
The critical points in the thermodynamic limit are
$\lambda_{c1}=2.51610\pm 3\times 10^{-5}$ and $\lambda_{c2}=2.53303\pm 4\times 10^{-5}$
for Large-$D$-ID and ID-Haldane, respectively.
The number of states kept is up to $m=1300$ in the largest size,
and the dimension is about $5\times 10^7$ in the Lanczos diagonalization of finding the lowest energy eigenstate.
The order of truncation error is $10^{-8}$ in the worst case.
In contrast, recent accurate determination of the Gaussian transition achieved the similar accuracy by
considering the entanglement entropy up to twenty thousand sites.\cite{Hu2011}
Therefore the LS method also becomes a very powerful tool for the Gaussian transitions.
In addition, following the LS method, it implies the presence of ID phase in model Eq.~(\ref{eq:s2xxzd}).
Thus Oshikawa's conjecture in 1992\cite{Oshikawa1992} is now supported by DMRG.
However, the region of the ID phase in $D=2.1$ is merely about $0.017$.
It may explain why the ID phase hasn't been found from previous DMRG studies
consequently.\cite{Schollwock1995,Schollwock1996,noID}

\begin{figure}[t]
\includegraphics[width=2.7in]{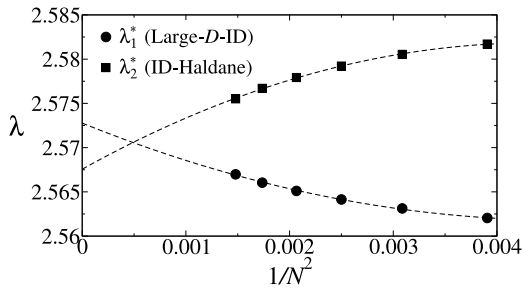}
\caption{$D=2.14$. Although the pseudo-critical points satisfy the condition Eq.~(\ref{eq:id}),
there is no energy level crossing in the thermodynamic limit.
In other words, there is no Gaussian transition.
This strong finite size effect is not observed\cite{Tonegawa2011}
until using the new parity DMRG method.}\label{fig:d214}
\end{figure}
\subsection{Intermediate-$D$ phase boundary}\label{sec:s2xxzd-b}
The phase diagram is focused on the region $D\geq 2.1$ which contains some ambiguity in previous studies.
A possible scenario is that the ID phase boundaries of ID-Haldane and Large-$D$-ID merge at a point.
This point $(\lambda_c,D_c)\simeq (2.64,2.19)$ was estimated by
extrapolation of phase boundaries from $D\lesssim 2$.\cite{Tonegawa2011}
In this scenario, the Haldane and Large-$D$ phases are suggested to be the same phase.\cite{Tonegawa2011,Pollmann2012}
However, from previous analysis by ED,\cite{Tonegawa2011}
it is known that larger value of $D$ leads the level crossing to begin at larger size.
Therefore, based on the presence of the ID phase, there are two other possible scenarios against the first one.
One is that the merge point $(\lambda_c,D_c)$ exactly locates at the boundary of
N\'eel phase and becomes a multi-critical point.
Another possible scenario is that the ID phase extends and reaches the N\'eel phase.
In the latter two scenarios, Haldane and Large-$D$ phases are
completely separated by the ID phase and definitely
not the same phase, though recent studies suggested
they could be the same phase.\cite{Tonegawa2011,Pollmann2012}

\begin{table}[b]
\caption{\label{tab:acc}%
The energies for $N=28$ in Eq.~(\ref{eq:id}) as $m$ increases.
$\lambda=2.51$ and $D=2.1$. Three sweeps are performed.}
\begin{ruledtabular}
\begin{tabular}{rccc}
$m$ & $p=+1$ & $p=-1$ & truncation error \\
\colrule
600   & -95.907406036 & -95.907356973 &$5.9\times 10^{-7}$ \\
700   & -95.907437223 & -95.907385487 &$2.9\times 10^{-7}$ \\
800   & -95.907451123 & -95.907399047 &$1.6\times 10^{-7}$ \\
900   & -95.907458008 & -95.907405896 &$9.6\times 10^{-8}$ \\
1000  & -95.907461893 & -95.907409748 &$5.8\times 10^{-8}$ \\
1100  & -95.907464072 & -95.907411938 &$3.7\times 10^{-8}$ \\
1200  & -95.907465368 & -95.907413199 &$2.5\times 10^{-8}$ \\
1300  & -95.907466194 & -95.907413994 &$1.7\times 10^{-8}$ \\
\end{tabular}
\end{ruledtabular}
\end{table}

\begin{figure}[t]
\includegraphics[width=2.7in]{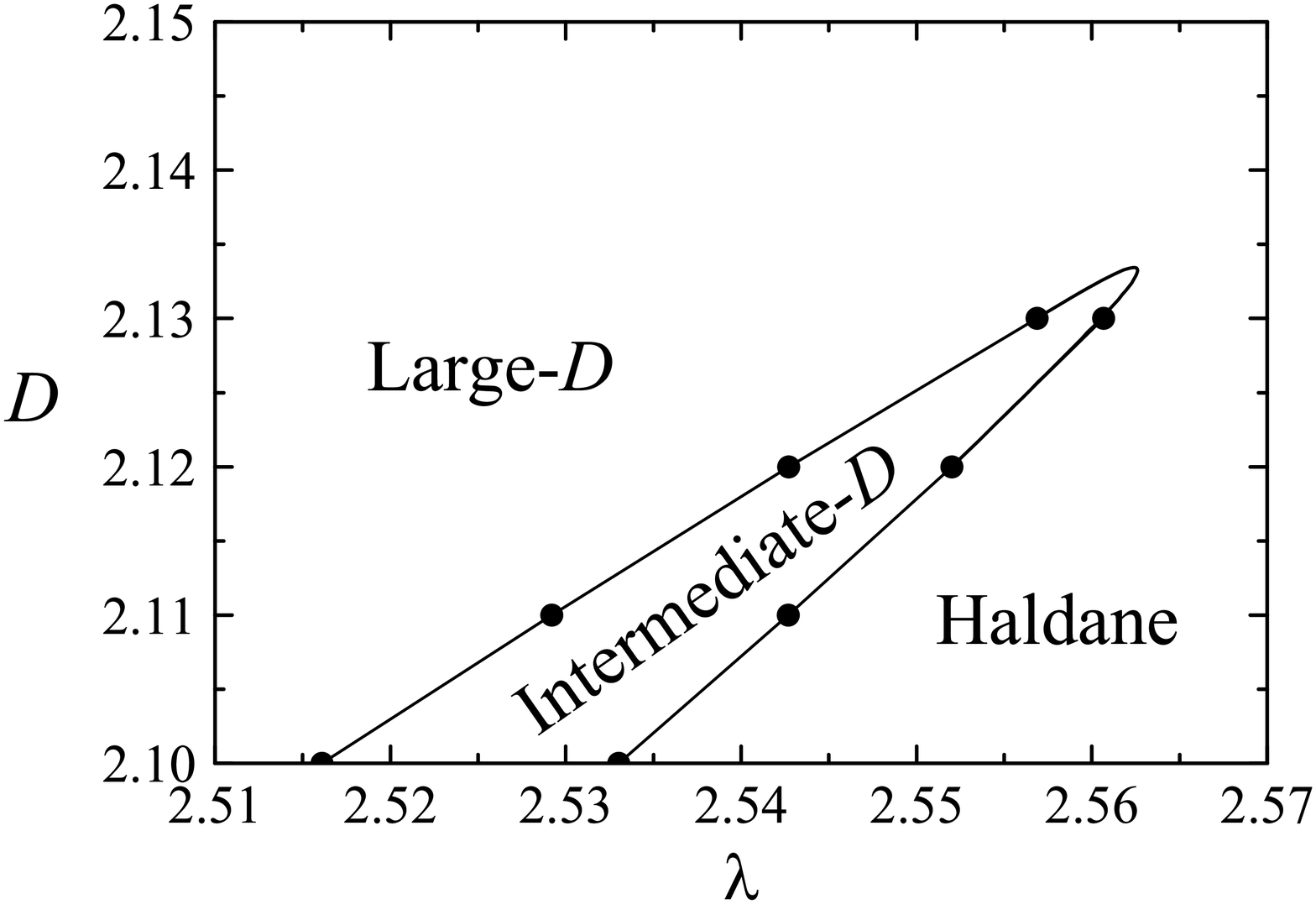}
\caption{The phase boundary of Intermediate-$D$ phase in the S=2 XXZ chain with uniaxial anisotropy.
the phase boundary lines between ID and Haldane phases
and between Large-$D$ and ID phases merge at a point between $2.13<D_c<2.14$.}\label{fig:idphase}
\end{figure}

Departing from empiric of ED, as shown in Fig.~\ref{fig:d214} for $D=2.14$, despite the pseudo-critical points satisfy
the condition Eq.~(\ref{eq:id}) in small sizes, there is no energy level crossing in the thermodynamic limit.
In other words, according to the LS method, there is no Gaussian transition, and
the scenarios that the ID phase completely separates Haldane and Large-$D$ phases are excluded.
The more precise phase diagram by means of the new parity DMRG is shown in Fig.~\ref{fig:idphase}.

Finally the accuracy for the current research is carefully examined.
It is known that the often-used measure of the error in DMRG calculations, the truncation error,
may not reveal the true relative error of energy in the \textit{infinite-system DMRG}.\cite{Sorensen1998}
In the \textit{finite-system DMRG}, the truncation error is a more reliable indication to the true error.
The energies of $\lambda=2.51$ and $D=2.1$ for $N=28$ in Eq.~(\ref{eq:id})
as the number of states kept $m$ increases are listed in Tab.~\ref{tab:acc}.
From these data, the absolute errors of energy in Fig.~\ref{fig:LS}(a)
are argued to be of order $10^{-6}$, and the relative errors to be the same order with the truncation error.
Therefore, the energies are sufficiently accurate for distinguishing the energy differences in the current studies.
The phase diagram in Fig.~\ref{fig:idphase} thus provides a solid numerical evidence
for the presence of the ID phase in the S=2 XXZ chain with uniaxial anisotropy Eq.~(\ref{eq:s2xxzd}).

\section{Conclusion}\label{sec:con}
The parity (space inversion) quantum numbers are of significance not only in
computational advances but also in the LS method. By employing the ladder scheme,
the global parity operator can be decomposed into the product form in Eq.~(\ref{eq:pi}).
It makes DMRG be able to utilize the parity quantum numbers.
The LS method is the first time performed by DMRG in the S=2 XXZ model with uniaxial anisotropy.
The BKT critical point of XY1-Large-$D$ as well as
Gaussain critical points of Large-$D$-ID and ID-Haldane are determined very precisely.
Thus the LS method is suggested to be the most powerful tool for detecting BKT and Gaussian transitions.

The phase diagram of the S=2 XXZ model with uniaxial anisotropy, Fig.~\ref{fig:idphase},
is investigated by focusing on the ID phase boundary
where the ED results are strongly affected by the finite size effect.
This work consistent with previous finding\cite{Tonegawa2011,Pollmann2012}
thus provides a first DMRG support to Oshikawa's conjecture\cite{Oshikawa1992} in 1992.

Since the proposed method in the case $d^2=25$ works well, it is also feasible
for a large number of models, including S=1/2 spin chains, S=1 spin chains, $t$-$J$ model, Hubbard model,
topological interacting fermion systems\cite{Guo2011}, and ladder, \textit{etc}.

\begin{figure}[t]
\includegraphics[width=2.7in]{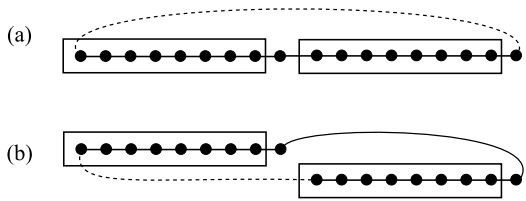}
\caption{The old parity schemes in the \textit{infinite-system DMRG}\cite{Sorensen1998}
for (a) PBC and (b) TBC. The dash line is the boundary coupling.
The left and right blocks must be the same length.
The truncation error may not reveal the true relative error of energy.\cite{Sorensen1998}
It is difficult to evaluate how large number of states kept $m$ is enough.
}\label{fig:sorensen}
\end{figure}
\begin{acknowledgments}
I am grateful to Ying-Jer Kao, Min-Fong Yang, Hsiang-Hsuan Hung,
Wei-Chao Shen, Hsiao-Ting Tzeng and Hui-Chen Tsai
for discussions and encouragement.
I am also grateful to Takashi Tonegawa for discussions about the scaling function.
The computations in this work were done on Acer AR585 F1 Cluster
of National Center for High-performance Computing, Taiwan.
This work is partially supported by Hsing Tian Kong Culture \& Education Development Foundation.
This work is supported by the NSC in Taiwan through Grants No. 100-2112-M-002-013-MY3.
\end{acknowledgments}
\begin{figure}[t]
\includegraphics[width=2.8in]{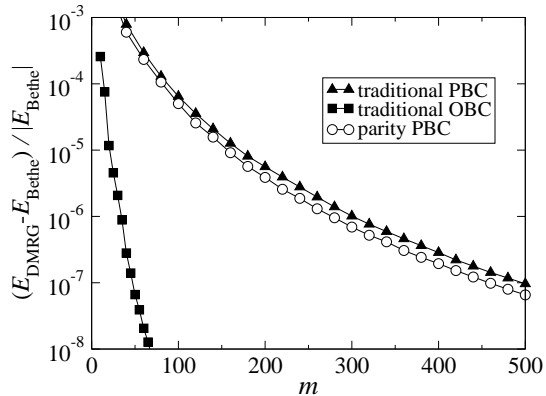}
\caption{Relative error in the ground state energy for $N=100$ sites spin-1/2 Heisenberg chain
as a function of the number of states kept $m$. $E_{\tiny{\mbox{DMRG}}}$ are computed by \textit{finite-system DMRG},
and the exact energy $E_{\tiny{\mbox{Bethe}}}$ are obtained from algebraic Bethe ansatz.\cite{Alcaraz1987}
}\label{fig:relative}
\end{figure}

\appendix*
\section{Boundary conditions and parity DMRG}
In the Level Spectroscopy (LS) method\cite{LS1997,LS1998},
Twisted Boundary Conditions (TBC) and parity quantum numbers are usually required in the numerical algorithm.
Although the old parity \textit{infinite-system DMRG}\cite{Sorensen1998} may obtain accurate energy within
Periodic Boundary Conditions (PBC) as long as enough large number of states are kept,
the PBC scheme in Fig.~\ref{fig:sorensen}(a) can not be used within TBC because
the parity described in Ref.~\onlinecite{Sorensen1998} is not conserved anymore.
Instead of Fig.~\ref{fig:sorensen}(a), the TBC scheme in Fig.~\ref{fig:sorensen}(b) should be used.
However it may not be a practical implementation because the direct connection between two blocks
slows down the algorithm dramatically.\cite{DMRG}
Moreover, the truncation error in the \textit{infinite-system DMRG}
may not reveal the true relative error of energy.\cite{Sorensen1998}
It is difficult to evaluate how large number of states kept $m$ is enough.
These considerations may be the reason why previous LS studies
were always performed by Exact Diagonalization.\cite{Nakamura1999,Chen2003,Hijii2005,Furukawa2010,Tonegawa2011}

The scheme in Fig.~\ref{fig:sorensen}(a) was first proposed by White in his initial DMRG papers for PBC.\cite{DMRG}
It is known that the traditional DMRG performs worse for PBC.
In order to improve the PBC case, the ladder scheme was first proposed by Qin \textit{et al.} in 1995.\cite{Qin1995}
Unfortunately, the comparison did not publish because the ladder scheme does not have a distinct improvement.
Fig.~\ref{fig:relative} shows the comparison of convergence with $m$ for \textit{finite-system DMRG}.
While the new parity DMRG within PBC has merely a little superiority to the traditional DMRG,
it should be emphasized that LS method which based on sine-Gordon theory has a strong numerical tool now.\cite{note}
For the PBC case, recent development by Pippan, White and Evertz have a remarkable improvement
with the Matrix Product State (MPS) algorithm.
Readers with interests are referred to Ref.~\onlinecite{Verstraete2004,Pippan2010} for more detail information.

%

\begin{thebibliography}{99}%
\makeatletter
\providecommand \@ifxundefined [1]{%
 \ifx #1\undefined \expandafter \@firstoftwo
 \else \expandafter \@secondoftwo
\fi
}%
\providecommand \@ifnum [1]{%
 \ifnum #1\expandafter \@firstoftwo
 \else \expandafter \@secondoftwo
\fi
}%
\providecommand \enquote [1]{``#1''}%
\providecommand \bibnamefont  [1]{#1}%
\providecommand \bibfnamefont [1]{#1}%
\providecommand \citenamefont [1]{#1}%
\providecommand\href[0]{\@sanitize\@href}%
\providecommand\@href[1]{\endgroup\@@startlink{#1}\endgroup\@@href}%
\providecommand\@@href[1]{#1\@@endlink}%
\providecommand \@sanitize [0]{\begingroup\catcode`\&12\catcode`\#12\relax}%
\@ifxundefined \pdfoutput {\@firstoftwo}{%
 \@ifnum{\z@=\pdfoutput}{\@firstoftwo}{\@secondoftwo}%
}{%
 \providecommand\@@startlink[1]{\leavevmode}%
 \providecommand\@@endlink[0]{}%
}{%
 \providecommand\@@startlink[1]{%
  \leavevmode
  \pdfstartlink
   attr{/Border[0 0 1 ]/H/I/C[0 1 1]}%
   user{/Subtype/Link/A<</Type/Action/S/URI/URI(#1)>>}%
  \relax
 }%
 \providecommand\@@endlink[0]{\pdfendlink}%
}%
\providecommand \url  [0]{\begingroup\@sanitize \@url }%
\providecommand \@url [1]{\endgroup\@href {#1}{\urlprefix}}%
\providecommand \urlprefix [0]{URL }%
\providecommand \Eprint[0]{\href }%
\@ifxundefined \urlstyle {%
  \providecommand \doi [1]{doi:\discretionary{}{}{}#1}%
}{%
  \providecommand \doi [0]{doi:\discretionary{}{}{}\begingroup
  \urlstyle{rm}\Url }%
}%
\providecommand \doibase [0]{http://dx.doi.org/}%
\providecommand \Doi[1]{\href{\doibase#1}}%
\providecommand \bibAnnote [3]{%
  \BibitemShut{#1}%
  \begin{quotation}\noindent
    \textsc{Key:}\ #2\\\textsc{Annotation:}\ #3%
  \end{quotation}%
}%
\providecommand \bibAnnoteFile [2]{%
  \IfFileExists{#2}{\bibAnnote {#1} {#2} {\input{#2}}}{}%
}%
\providecommand \typeout [0]{\immediate \write \m@ne }%
\providecommand \selectlanguage [0]{\@gobble}%
\providecommand \bibinfo [0]{\@secondoftwo}%
\providecommand \bibfield [0]{\@secondoftwo}%
\providecommand \translation [1]{[#1]}%
\providecommand \BibitemOpen[0]{}%
\providecommand \bibitemStop [0]{}%
\providecommand \bibitemNoStop [0]{.\EOS\space}%
\providecommand \EOS [0]{\spacefactor3000\relax}%
\providecommand \BibitemShut [1]{\csname bibitem#1\endcsname}%
\bibitem{BKT}%
  \BibitemOpen
  \bibfield{author}{\bibinfo {author} {\bibfnamefont{V.~L.}\ \bibnamefont{Berezinskii}}},
  \bibfield{journal}{Zh. Eksp. Teor. Fiz.}\ \textbf{\bibinfo {volume} {61}},\ %
  \bibinfo {pages} {1144} (\bibinfo {year}{1971});
  \bibfield{author}{\bibinfo {author} {\bibfnamefont{J.~M.}\ \bibnamefont{Kosterlitz}}\ and\ \bibinfo
  {author} {\bibfnamefont{D.~J.}\ \bibnamefont{Thouless}},\ }%
  \bibfield{journal}{\Doi{10.1088/0022-3719/6/7/010}{\bibinfo {journal} {J. Phys. C}}}
  \textbf{\bibinfo {volume} {6}},\ \bibinfo {pages} {1181} (\bibinfo {year}{1973})
  \bibAnnoteFile{NoStop}{BKT}%
\bibitem{BKTmethod}%
  \BibitemOpen
  \bibfield{author}{\bibinfo {author} {\bibfnamefont{L.}~\bibnamefont{Campos~Venuti}}, \bibinfo
  {author} {\bibfnamefont{C.}~\bibnamefont{Degli Esposti~Boschi}}, \bibinfo
  {author} {\bibfnamefont{M.}~\bibnamefont{Roncaglia}},\ and\ \bibinfo {author}
  {\bibfnamefont{A.}~\bibnamefont{Scaramucci}},\ }%
  \bibfield{journal}{\Doi{10.1103/PhysRevA.73.010303}{\bibinfo {journal} {\pra}}\ }%
  \textbf{\bibinfo {volume} {73}},\ \bibinfo {pages} {010303}(R) (\bibinfo {year}{2006});
  \bibfield{author}{\bibinfo {author} {\bibfnamefont{M.}~\bibnamefont{Roncaglia}}, \bibinfo
  {author} {\bibfnamefont{L.}~\bibnamefont{Campos~Venuti}},\ and\ \bibinfo
  {author} {\bibfnamefont{C.}~\bibnamefont{Degli Esposti~Boschi}},\ }%
  \bibfield{journal}{\Doi{10.1103/PhysRevB.77.155413}{\bibinfo {journal} {\prb}}\ }%
  \textbf{\bibinfo {volume} {77}},\ \bibinfo {pages} {155413} (\bibinfo {year}{2008})%
  \bibAnnoteFile{NoStop}{BKTmethod}%
\bibitem{Zhou}%
  \BibitemOpen
  \bibfield{author}{\bibinfo {author} {\bibfnamefont{H.-L.}~\bibnamefont{Wang}}, \bibinfo
  {author} {\bibfnamefont{A.-M.}~\bibnamefont{Chen}}, \bibinfo
  {author} {\bibfnamefont{B.}~\bibnamefont{Li}},\ and\ \bibinfo {author}
  {\bibfnamefont{H.-Q.}~\bibnamefont{Zhou}},\ }%
  \bibfield{journal}{\Doi{10.1088/1751-8113/45/1/015306}{\bibinfo {journal} {J. Phys. A}}\ }%
  \textbf{\bibinfo {volume} {45}},\ \bibinfo {pages} {015306} (\bibinfo {year}{2012});
  \bibfield{author}{\bibinfo {author} {\bibfnamefont{H.-L.}~\bibnamefont{Wang}}, \bibinfo
  {author} {\bibfnamefont{J.-H.}~\bibnamefont{Zhao}}, \bibinfo
  {author} {\bibfnamefont{B.}~\bibnamefont{Li}},\ and\ \bibinfo {author}
  {\bibfnamefont{H.-Q.}~\bibnamefont{Zhou}},\ }%
  \bibfield{journal}{\Doi{10.1088/1742-5468/2011/10/L10001}{\bibinfo {journal} {J. Stat. Mech.: Theor. Exp.}}\ }%
  (\textbf{\bibinfo{volume}{2011}}),\ \bibinfo {pages} {L10001};
  \bibfield{author}{\bibinfo {author} {\bibfnamefont{B.}~\bibnamefont{Wang}}, \bibinfo
  {author} {\bibfnamefont{M.}~\bibnamefont{Feng}},\ and\ \bibinfo {author}
  {\bibfnamefont{Z.-Q.}~\bibnamefont{Chen}},\ }%
  \bibfield{journal}{\Doi{10.1103/PhysRevA.81.064301}{\bibinfo {journal} {\pra}}\ }%
  \textbf{\bibinfo {volume} {81}},\ \bibinfo {pages} {064301} (\bibinfo {year}{2010})
\bibAnnoteFile{NoStop}{Zhou}%
\bibitem{MFYang}%
  \BibitemOpen
  \bibfield{author}{\bibinfo {author} {\bibfnamefont{M.-F.}\ \bibnamefont{Yang}},\ }%
  \bibfield{journal}{\Doi{10.1103/PhysRevA.71.030302}{\bibinfo {journal} {\pra}}\ }%
  \textbf{\bibinfo {volume} {71}},\ \bibinfo {pages} {030302(R)} (\bibinfo {year}{2005});
  \bibfield{author}{\bibinfo {author} {\bibfnamefont{M.-F.}\ \bibnamefont{Yang}},\ }%
  \bibfield{journal}{\Doi{10.1103/PhysRevB.76.180403}{\bibinfo {journal} {\prb}}\ }%
  \textbf{\bibinfo {volume} {76}},\ \bibinfo {pages} {180403(R)} (\bibinfo {year}{2007})%
  \bibAnnoteFile{NoStop}{MFYang}%
\bibitem{Tzeng}%
  \BibitemOpen
  \bibfield{author}{\bibinfo {author} {\bibfnamefont{Y.-C.}\ \bibnamefont{Tzeng}}\ and\ \bibinfo
  {author} {\bibfnamefont{M.-F.}\ \bibnamefont{Yang}},\ }%
  \bibfield{journal}{\Doi{10.1103/PhysRevA.77.012311}{\bibinfo {journal} {\pra}}\ }%
  \textbf{\bibinfo {volume} {77}},\ \bibinfo {pages} {012311} (\bibinfo {year}{2008});
  \bibfield{author}{\bibinfo {author} {\bibfnamefont{Y.-C.}\ \bibnamefont{Tzeng}}, \bibinfo
  {author} {\bibfnamefont{H.-H.}\ \bibnamefont{Hung}}, \bibinfo {author}
  {\bibfnamefont{Y.-C.}\ \bibnamefont{Chen}},\ and\ \bibinfo {author}
  {\bibfnamefont{M.-F.}\ \bibnamefont{Yang}},\ }%
  \bibfield{journal}{\Doi{10.1103/PhysRevA.77.062321}{\bibinfo {journal} {Phys. Rev. A}}\ }%
  \textbf{\bibinfo {volume} {77}},\ \bibinfo {pages} {062321} (\bibinfo {year}{2008});
  \bibfield{author}{\bibinfo {author} {\bibfnamefont{M.}\ \bibnamefont{Thesberg}}\ and\ \bibinfo
  {author} {\bibfnamefont{E.~S.}\ \bibnamefont{S\o{}rensen}},\ }%
  \bibfield{journal}{\Doi{10.1103/PhysRevB.84.224435}{\bibinfo {journal} {\prb}}\ }%
  \textbf{\bibinfo {volume} {84}},\ \bibinfo {pages} {224435} (\bibinfo {year}{2011})
  \bibAnnoteFile{NoStop}{Tzeng}%
\bibitem{Hu2011}%
  \BibitemOpen
  \bibfield{author}{\bibinfo {author} {\bibfnamefont{S.}~\bibnamefont{Hu}}, \bibinfo {author}
  {\bibfnamefont{B.}~\bibnamefont{Normand}}, \bibinfo {author}
  {\bibfnamefont{X.}~\bibnamefont{Wang}},\ and\ \bibinfo {author}
  {\bibfnamefont{L.}~\bibnamefont{Yu}},\ }%
  \bibfield{journal}{\Doi{10.1103/PhysRevB.84.220402}{\bibinfo {journal} {\prb}}\ }%
  \textbf{\bibinfo {volume} {84}},\ \bibinfo {pages} {220402}(R) (\bibinfo {year}{2011})%
  \bibAnnoteFile{NoStop}{Hu2011}%
\bibitem{Rachel2012}%
  \BibitemOpen
  \bibfield{author}{\bibinfo {author} {\bibfnamefont{S.}~\bibnamefont{Rachel}}, \bibinfo {author}
  {\bibfnamefont{N.}~\bibnamefont{Laflorencie}}, \bibinfo {author}
  {\bibfnamefont{H.~F.}\ \bibnamefont{Song}},\ and\ \bibinfo {author}
  {\bibfnamefont{K.}~\bibnamefont{Le~Hur}},\ }%
  \bibfield{journal}{\Doi{10.1103/PhysRevLett.108.116401}{\bibinfo {journal} {\prl}}\ }%
  \textbf{\bibinfo {volume} {108}},\ \bibinfo {pages} {116401} (\bibinfo {year}{2012})%
  \bibAnnoteFile{NoStop}{Rachel2012}%
\bibitem{LS1997}%
  \BibitemOpen
  \bibfield{author}{\bibinfo {author} {\bibfnamefont{K.}~\bibnamefont{Nomura}},\ }%
  \bibfield{journal}{\Doi{10.1088/0305-4470/28/19/003}{\bibinfo {journal} {J. Phys. A}}\ }%
  \textbf{\bibinfo {volume} {28}},\ \bibinfo {pages} {5451} (\bibinfo {year}{1995});
  \bibfield{author}{\bibinfo {author} {\bibfnamefont{A.}~\bibnamefont{Kitazawa}},\ }%
  \bibfield{journal}{\Doi{10.1088/0305-4470/30/9/005}{\bibinfo {journal} {J. Phys. A }}}%
  \textbf{\bibinfo {volume} {30}},\ \bibinfo {pages} {L285} (\bibinfo {year}{1997})
  \bibAnnoteFile{NoStop}{LS1997}%
\bibitem{LS1998}%
  \BibitemOpen 
  \bibfield{author}{\bibinfo {author} {\bibfnamefont{K.}~\bibnamefont{Nomura}}\ and\ \bibinfo
  {author} {\bibfnamefont{A.}~\bibnamefont{Kitazawa}},\ }%
  \bibfield{journal}{\Doi{10.1088/0305-4470/31/36/008}{\bibinfo {journal} {J. Phys. A}}\ }%
  \textbf{\bibinfo {volume} {31}},\ \bibinfo {pages} {7341} (\bibinfo {year}{1998})%
  \bibAnnoteFile{NoStop}{LS1998}%
\bibitem{Nakamura1999}%
  \BibitemOpen
  \bibfield{author}{\bibinfo {author} {\bibfnamefont{M.}~\bibnamefont{Nakamura}},\ }%
  \bibfield{journal}{\Doi{10.1143/JPSJ.68.3123}{\bibinfo {journal} {J. Phys. Soc. Jpn.}}\ }%
  \textbf{\bibinfo {volume} {68}},\ \bibinfo {pages} {3123} (\bibinfo {year}{1999});
  \bibfield{author}{\bibinfo {author} {\bibfnamefont{M.}~\bibnamefont{Nakamura}},\ }%
  \bibfield{journal}{\Doi{10.1103/PhysRevB.61.16377}{\bibinfo {journal} {Phys. Rev. B}}\ }%
  \textbf{\bibinfo {volume} {61}},\ \bibinfo {pages} {16377} (\bibinfo {year}{2000});
  \bibfield{author}{\bibinfo {author} {\bibfnamefont{H.}~\bibnamefont{Otsuka}}\ and\ \bibinfo
  {author} {\bibfnamefont{M.}~\bibnamefont{Nakamura}},\ }%
  \bibfield{journal}{\Doi{10.1103/PhysRevB.71.155105}{\bibinfo {journal} {\prb}}}
  \textbf{\bibinfo {volume} {71}},\ \bibinfo {pages} {155105} (\bibinfo {year}{2005})%
  \bibAnnoteFile{NoStop}{Nakamura1999}%
\bibitem{Chen2003}%
  \BibitemOpen
  \bibfield{author}{\bibinfo {author} {\bibfnamefont{W.}~\bibnamefont{Chen}}, \bibinfo {author}
  {\bibfnamefont{K.}~\bibnamefont{Hida}},\ and\ \bibinfo {author}
  {\bibfnamefont{B.~C.}\ \bibnamefont{Sanctuary}},\ }%
  \bibfield{journal}{\Doi{10.1143/JPSJ.69.237}{\bibinfo {journal} {J. Phys. Soc. Jpn.}}\ }%
  \textbf{\bibinfo {volume} {69}},\ \bibinfo {pages} {237} (\bibinfo {year}{2000});
  \bibfield{author}{\bibinfo {author} {\bibfnamefont{W.}~\bibnamefont{Chen}}, \bibinfo {author}
  {\bibfnamefont{K.}~\bibnamefont{Hida}},\ and\ \bibinfo {author}
  {\bibfnamefont{B.~C.}\ \bibnamefont{Sanctuary}},\ }%
  \bibfield{journal}{\Doi{10.1103/PhysRevB.67.104401}{\bibinfo {journal} {Phys. Rev. B}}\ }%
  \textbf{\bibinfo {volume} {67}},\ \bibinfo {pages} {104401} (\bibinfo {year}{2003});
  \bibfield{author}{\bibinfo {author} {\bibfnamefont{T.}~\bibnamefont{Murashima}}, \bibinfo {author}
  {\bibfnamefont{K.}~\bibnamefont{Hijii}}, \bibinfo {author}{\bibfnamefont{K.}~\bibnamefont{Nomura}},
  and\ \bibinfo {author} {\bibfnamefont{T.}\ \bibnamefont{Tonegawa}},\ }%
  \bibfield{journal}{\Doi{10.1143/JPSJ.74.1544}{\bibinfo {journal} {J. Phys. Soc. Jpn.}}\ }%
  \textbf{\bibinfo {volume} {74}},\ \bibinfo {pages} {1544} (\bibinfo {year}{2005}) 
  \bibAnnoteFile{NoStop}{Chen2003}%
\bibitem{Hijii2005}%
  \BibitemOpen
  \bibfield{author}{\bibinfo {author} {\bibfnamefont{K.}~\bibnamefont{Hijii}}, \bibinfo {author}%
  {\bibfnamefont{A.}~\bibnamefont{Kitazawa}},\ and\ \bibinfo {author}%
  {\bibfnamefont{K.}\ \bibnamefont{Nomura}},\ }%
  \bibfield{journal}{\Doi{10.1103/PhysRevB.72.014449}{\bibinfo {journal} {Phys. Rev. B}}\ }%
  \textbf{\bibinfo {volume} {72}},\ \bibinfo {pages} {014449} (\bibinfo {year}{2005});
  \bibfield{author}{\bibinfo {author} {\bibfnamefont{M.}~\bibnamefont{Nakamura}},\ }%
  \bibfield{journal}{\Doi{10.1016/S0921-4526(02)02180-4}{\bibinfo {journal} {Physica B}}\ }%
  \textbf{\bibinfo {volume} {329}},\ \bibinfo {pages} {1000} (\bibinfo {year}{2003});
  \bibfield{author}{\bibinfo {author} {\bibfnamefont{M.}~\bibnamefont{Nakamura}}, \bibinfo {author}
  {\bibfnamefont{T.}~\bibnamefont{Yamamoto}},\ and\ \bibinfo {author}
  {\bibfnamefont{K.}\ \bibnamefont{Ide}},\ }%
  \bibfield{journal}{\Doi{10.1143/JPSJ.72.1022}{\bibinfo {journal} {J. Phys. Soc. Jpn.}}\ }%
  \textbf{\bibinfo {volume} {72}},\ \bibinfo {pages} {1022} (\bibinfo {year}{2003});
  \bibfield{author}{\bibinfo {author} {\bibfnamefont{H.}~\bibnamefont{Otsuka}}, \bibinfo {author}
  {\bibfnamefont{Y.}~\bibnamefont{Okabe}},\ and\ \bibinfo {author}
  {\bibfnamefont{K.}\ \bibnamefont{Okunishi}},\ }%
  \bibfield{journal}{\Doi{10.1103/PhysRevE.73.035105}{\bibinfo {journal} {Phys. Rev. E}}\ }%
  \textbf{\bibinfo {volume} {73}},\ \bibinfo {pages} {035105}(R) (\bibinfo {year}{2006})%
  \bibAnnoteFile{NoStop}{Hijii2005}%
\bibitem{Furukawa2010}%
  \BibitemOpen
  \bibfield{author}{\bibinfo {author} {\bibfnamefont{S.}~\bibnamefont{Furukawa}}, \bibinfo
  {author} {\bibfnamefont{M.}~\bibnamefont{Sato}},\ and\ \bibinfo {author}
  {\bibfnamefont{A.}~\bibnamefont{Furusaki}},\ }%
  \bibfield{journal}{\Doi{10.1103/PhysRevB.81.094430}{\bibinfo {journal} {Phys. Rev. B}}\ }%
  \textbf{\bibinfo {volume} {81}},\ \bibinfo {pages} {094430} (\bibinfo {year}{2010})%
  \bibAnnoteFile{NoStop}{Furukawa2010}%
\bibitem{Tonegawa2011}%
  \BibitemOpen
  \bibfield{author}{\bibinfo {author} {\bibfnamefont{T.}~\bibnamefont{Tonegawa}}, \bibinfo
  {author} {\bibfnamefont{K.}~\bibnamefont{Okamoto}}, \bibinfo {author}
  {\bibfnamefont{H.}~\bibnamefont{Nakano}}, \bibinfo {author}
  {\bibfnamefont{T.}~\bibnamefont{Sakai}}, \bibinfo {author}
  {\bibfnamefont{K.}~\bibnamefont{Nomura}},\ and\ \bibinfo {author}
  {\bibfnamefont{M.}~\bibnamefont{Kaburagi}},\ }%
  \bibfield{journal}{\Doi{10.1143/JPSJ.80.043001}{\bibinfo {journal} {J. Phys. Soc. Jpn.}}\ }%
  \textbf{\bibinfo {volume} {80}},\ \bibinfo {pages} {043001} (\bibinfo {year}{2011})%
  \bibAnnoteFile{NoStop}{Tonegawa2011}%
\bibitem{HQLin1990}%
  \BibitemOpen
  \bibfield{author}{\bibinfo {author} {\bibfnamefont{H.-Q.}\ \bibnamefont{Lin}},\ }%
  \bibfield{journal}{\Doi{10.1103/PhysRevB.42.6561}{\bibinfo {journal} {Phys. Rev. B}}\ }%
  \textbf{\bibinfo {volume} {42}},\ \bibinfo {pages} {6561} (\bibinfo {year}{1990})%
  \bibAnnoteFile{NoStop}{HQLin1990}%
\bibitem{Guo2011}%
  \BibitemOpen
  \bibfield{author}{\bibinfo {author} {\bibfnamefont{H.}~\bibnamefont{Guo}}\ and\ \bibinfo
  {author} {\bibfnamefont{S.-Q.}\ \bibnamefont{Shen}},\ }%
  \bibfield{journal}{\Doi{10.1103/PhysRevB.84.195107}{\bibinfo {journal} {Phys. Rev. B}}\ }%
  \textbf{\bibinfo {volume} {84}},\ \bibinfo {pages} {195107} (\bibinfo {year} {2011})%
  \bibAnnoteFile{NoStop}{Guo2011}%
\bibitem{SPT}%
  \BibitemOpen
  \bibinfo {note} {Such an algorithm is also helpful for recent development of
  classification of Symmetry Protected Topological phases in 1D, see}
  \bibfield{author}{\bibinfo {author} {\bibfnamefont{X.}~\bibnamefont{Chen}}, \bibinfo {author}%
  {\bibfnamefont{Z.-C.}~\bibnamefont{Gu}}\ and\ \bibinfo {author}%
  {\bibfnamefont{X.-G.}~\bibnamefont{Wen}},\ }%
  \bibfield{journal}{\Doi{10.1103/PhysRevB.83.035107}{\bibinfo {journal} {Phys. Rev. B}}\ }%
  \textbf{\bibinfo {volume} {83}},\ \bibinfo {pages} {035107} (\bibinfo {year} {2011});
  \bibfield{journal}{\Doi{10.1103/PhysRevB.84.235128}{\bibinfo {journal} {\prb}}\ }%
  \textbf{\bibinfo {volume} {84}},\ \bibinfo {pages} {235128} (\bibinfo {year} {2011});
  \bibfield{author}{\bibinfo {author} {\bibfnamefont{X.-G.}~\bibnamefont{Wen}},\ }%
  \bibfield{journal}{\Doi{10.1103/PhysRevB.85.085103}{\bibinfo {journal} {\prb}}}
  \textbf{\bibinfo {volume} {85}},\ \bibinfo {pages} {085103} (\bibinfo {year} {2012});
  \bibfield{author}{\bibinfo {author} {\bibfnamefont{Z.-C.}~\bibnamefont{Gu}}\ and\ \bibinfo {author}%
  {\bibfnamefont{X.-G.}~\bibnamefont{Wen}},\ }%
  \bibfield{journal}{\Doi{10.1103/PhysRevB.80.155131}{\bibinfo {journal} {Phys. Rev. B\ }}}%
  \textbf{\bibinfo {volume} {80}},\ \bibinfo {pages} {155131} (\bibinfo {year} {2009});
  \Eprint{http://arxiv.org/abs/1201.2648}{arXiv:1201.2648};
  \bibfield{author}{\bibinfo {author} {\bibfnamefont{E.}~\bibnamefont{Tang}}\ and\ \bibinfo {author}%
  {\bibfnamefont{X.-G.}~\bibnamefont{Wen}},\ }\Eprint{http://arxiv.org/abs/1204.0520}{arXiv:1204.0520};
  \bibfield{author}{\bibinfo {author} {\bibfnamefont{F.}~\bibnamefont{Pollmann}}, \bibinfo {author}%
  {\bibfnamefont{A.~M.}~\bibnamefont{Turner}}, \bibinfo {author}%
  {\bibfnamefont{E.}~\bibnamefont{Berg}},\ and\ \bibinfo {author}%
  {\bibfnamefont{M.}~\bibnamefont{Oshikawa}},\ }%
  \bibfield{journal}{\Doi{10.1103/PhysRevB.81.064439}{\bibinfo {journal} {Phys. Rev. B}}\ }%
  \textbf{\bibinfo {volume} {81}},\ \bibinfo {pages} {064439} (\bibinfo {year} {2010});
  \bibfield{author}{\bibinfo {author} {\bibfnamefont{F.}~\bibnamefont{Pollmann}}\ and\ \bibinfo {author}%
  {\bibfnamefont{A. M.}~\bibnamefont{Turner}},\ }\Eprint{http://arxiv.org/abs/1204.0704}{arXiv:1204.0704}
  \bibAnnoteFile{NoStop}{SPT}%
\bibitem{DMRG}%
  \BibitemOpen
  \bibfield{author}{\bibinfo {author} {\bibfnamefont{S.~R.}\ \bibnamefont{White}},\ }%
  \bibfield{journal}{\Doi{10.1103/PhysRevLett.69.2863}{\bibinfo {journal} {Phys. Rev. Lett.}}\ }%
  \textbf{\bibinfo {volume} {69}},\ \bibinfo {pages} {2863} (\bibinfo {year}{1992});
  \bibfield{author}{\bibinfo {author} {\bibfnamefont{S.~R.}\ \bibnamefont{White}},\ }%
  \bibfield{journal}{\Doi{10.1103/PhysRevB.48.10345}{\bibinfo {journal} {Phys. Rev. B}}\ }%
  \textbf{\bibinfo {volume} {48}},\ \bibinfo {pages} {10345} (\bibinfo {year} {1993})%
  \bibAnnoteFile{NoStop}{DMRG}%
\bibitem{White1996}%
  \BibitemOpen
  \bibfield{author}{\bibinfo {author} {\bibfnamefont{S.~R.}\ \bibnamefont{White}},\ }%
  \bibfield{journal}{\Doi{10.1103/PhysRevLett.77.3633}{\bibinfo {journal} {Phys. Rev. Lett.}}\ }%
  \textbf{\bibinfo {volume} {77}},\ \bibinfo {pages} {3633} (\bibinfo {year} {1996})%
  \bibAnnoteFile{NoStop}{White1996}%
\bibitem{White2005}%
  \BibitemOpen
  \bibfield{author}{\bibinfo {author} {\bibfnamefont{S.~R.}\ \bibnamefont{White}},\ }%
  \bibfield{journal}{\Doi{10.1103/PhysRevB.72.180403}{\bibinfo {journal} {Phys. Rev. B}}\ }%
  \textbf{\bibinfo {volume} {72}},\ \bibinfo {pages} {180403}(R) (\bibinfo {year} {2005})
  \bibAnnoteFile{NoStop}{White2005}%
\bibitem{RMP2005}%
  \BibitemOpen
  \bibfield{author}{\bibinfo {author} {\bibfnamefont{U.}~\bibnamefont{Schollw\"ock}},\ }%
  \bibfield{journal}{\Doi{10.1103/RevModPhys.77.259}{\bibinfo {journal} {\rmp}}\ }%
  \textbf{\bibinfo {volume} {77}},\ \bibinfo {pages} {259} (\bibinfo {year}{2005});
  \bibfield{author}{\bibinfo {author} {\bibfnamefont{K.~A.}~\bibnamefont{Hallberg}},\ }%
  \bibfield{journal}{\Doi{10.1080/00018730600766432}{\bibinfo {journal} {Adv. Phys.}}\ }%
  \textbf{\bibinfo {volume} {55}},\ \bibinfo {pages} {477} (\bibinfo {year}{2006})%
  \bibAnnoteFile{NoStop}{RMP2005}%
\bibitem{Yan2011}%
  \BibitemOpen
  \bibfield{author}{\bibinfo {author} {\bibfnamefont{S.}~\bibnamefont{Yan}}, \bibinfo {author}
  {\bibfnamefont{D.~A.}\ \bibnamefont{Huse}},\ and\ \bibinfo {author}
  {\bibfnamefont{S.~R.}\ \bibnamefont{White}},\ }%
  \bibfield{journal}{\Doi{10.1126/science.1201080}{\bibinfo {journal} {Science}}\ }%
  \textbf{\bibinfo {volume} {332}},\ \bibinfo {pages} {1173} (\bibinfo {year} {2011})%
  \bibAnnoteFile{NoStop}{Yan2011}%
\bibitem{McCulloch}%
  \BibitemOpen
  \bibfield{author}{\bibinfo {author} {\bibfnamefont{I.~P.}\ \bibnamefont{McCulloch}}\ and\
  \bibinfo {author} {\bibfnamefont{M.}~\bibnamefont{Gul\'acsi}},\ }%
  \bibfield{journal}{\bibinfo{journal} {Aust. J. Phys.}\ }%
  \textbf{\bibinfo {volume} {53}},\ \bibinfo {pages} {4} (\bibinfo {year}{2000});
  \bibfield{journal}{\bibinfo{journal} {Phil. Mag. Lett.}\ }%
  \textbf{\bibinfo {volume} {81}},\ \bibinfo {pages} {447} (\bibinfo {year}{2001});
  \bibfield{journal}{\Doi{10.1209/epl/i2002-00393-0}{\bibinfo{journal} {Europhys. Lett.}}\ }%
  \textbf{\bibinfo {volume} {57}},\ \bibinfo {pages} {852} (\bibinfo {year}{2002});
  \bibfield{author}{\bibinfo {author} {\bibfnamefont{D.}\ \bibnamefont{Zgid}}\ and\
  \bibinfo {author} {\bibfnamefont{M.}~\bibnamefont{Nooijen}},\ }%
  \bibfield{journal}{\Doi{10.1063/1.2814150}{\bibinfo{journal} {J. Chem. Phys.}}\ }%
  \textbf{\bibinfo {volume} {128}},\ \bibinfo {pages} {014107} (\bibinfo {year}{2008})%
  \bibAnnoteFile{NoStop}{McCulloch}%
\bibitem{Sorensen1998}%
  \BibitemOpen
  \bibfield{author}{\bibinfo {author} {\bibfnamefont{E.~S.}\ \bibnamefont{S\o{}rensen}},\ }%
  \bibfield{journal}{\Doi{10.1088/0953-8984/10/47/016}{\bibinfo {journal} {J. Phys.: Cond. Matt.}}\ }%
  \textbf{\bibinfo {volume} {10}},\ \bibinfo {pages} {10655} (\bibinfo {year} {1998})%
  \bibAnnoteFile{NoStop}{Sorensen1998}%
\bibitem{Ramasesha1996}%
  \BibitemOpen
  \bibfield{author}{\bibinfo {author} {\bibfnamefont{S.}~\bibnamefont{Ramasesha}}, \bibinfo {author}
  {\bibfnamefont{S.~K.}~\bibnamefont{Pati}}, \bibinfo {author}
  {\bibfnamefont{H.~R.}~\bibnamefont{Krishnamurthy}}, \bibinfo {author}
  {\bibfnamefont{Z.}~\bibnamefont{Shuai}},\ and\ \bibinfo {author}
  {\bibfnamefont{J.~L.}~\bibnamefont{Br\'edas}},\ }%
  \bibfield{journal}{\Doi{10.1103/PhysRevB.54.7598}{\bibinfo {journal} {\prb}}\ }%
  \textbf{\bibinfo {volume} {54}},\ \bibinfo {pages} {7598} (\bibinfo {year}{1996})%
  \bibAnnoteFile{NoStop}{Ramasesha1996}%
\bibitem{Rommer1997}%
  \BibitemOpen
  \bibfield{author}{\bibinfo {author} {\bibfnamefont{S.}~\bibnamefont{Rommer}}\ and\ \bibinfo
  {author} {\bibfnamefont{S.}~\bibnamefont{\"Ostlund}},\ }%
  \bibfield{journal}{\Doi{10.1103/PhysRevB.55.2164}{\bibinfo {journal} {Phys. Rev. B}}\ }%
  \textbf{\bibinfo {volume} {55}},\ \bibinfo {pages} {2164} (\bibinfo {year} {1997})%
  \bibAnnoteFile{NoStop}{Rommer1997}%
\bibitem{MPS}%
  \BibitemOpen
  \bibfield{author}{\bibinfo {author} {\bibfnamefont{I.~P.}~\bibfnamefont{McCulloch}},\ }%
  \bibfield{journal}{\Doi{10.1088/1742-5468/2007/10/P10014}{\bibinfo {journal}{J. Stat. Mech.: Theor. Exp.}}\ }%
  (\textbf{\bibinfo{volume}{2007}})\ \bibinfo {pages} {P10014};
  \bibfield{author}{\bibinfo {author} {\bibfnamefont{U.}~\bibfnamefont{Schollw\"ock}},\ }%
  \bibfield{journal}{\Doi{10.1016/j.aop.2010.09.012}{\bibinfo {journal}{Ann. Phys.}}\ }%
  \textbf{\bibinfo{volume}{326}},\ \bibinfo {pages} {96} (\bibinfo {year}{2011})%
  \bibAnnoteFile{NoStop}{MPS}%
\bibitem{good}%
  \BibitemOpen
  \bibfield{author}{\bibinfo {author} {\bibfnamefont{A.}~\bibnamefont{Weichselbaum}}\ and\
  \bibinfo {author} {\bibfnamefont{S.~R.}\ \bibnamefont{White}},\ }%
  \bibfield{journal}{\Doi{10.1103/PhysRevB.84.245130}{\bibinfo {journal} {Phys. Rev. B}}\ }%
  \textbf{\bibinfo {volume} {84}},\ \bibinfo {pages} {245130} (\bibinfo {year} {2011})%
  \bibAnnoteFile{NoStop}{good}%
\bibitem{Oshikawa1992}%
  \BibitemOpen
  \bibfield{author}{\bibinfo {author} {\bibfnamefont{M.}~\bibnamefont{Oshikawa}},\ }%
  \bibfield{journal}{\Doi{10.1088/0953-8984/4/36/019}{\bibinfo {journal} {J. Phys.: Cond. Matt.}}\ }%
  \textbf{\bibinfo {volume} {4}},\ \bibinfo {pages} {7469} (\bibinfo {year}{1992})%
  \bibAnnoteFile{NoStop}{Oshikawa1992}%
\bibitem{CNYang1966}%
  \BibitemOpen
  \bibfield{author}{\bibinfo {author} {\bibfnamefont{C.~N.}\ \bibnamefont{Yang}}\ and\ \bibinfo
  {author} {\bibfnamefont{C.~P.}\ \bibnamefont{Yang}},\ }%
  \bibfield{journal}{\Doi{10.1103/PhysRev.150.321}{\bibinfo {journal} {Phys. Rev.}}\ }%
  \textbf{\bibinfo {volume} {150}},\ \bibinfo {pages} {321} (\bibinfo {year} {1966});
  \bibfield{author}{\bibinfo {author} {\bibfnamefont{M.}\ \bibnamefont{Gaudin}},\ }%
  \bibfield{journal}{\Doi{10.1103/PhysRevA.4.386}{\bibinfo {journal} {Phys. Rev. A}}\ }%
  \textbf{\bibinfo {volume} {4}},\ \bibinfo {pages} {386} (\bibinfo {year} {1971})
  \bibAnnoteFile{NoStop}{CNYang1966}%
\bibitem{Alcaraz1987}%
  \BibitemOpen
  \bibfield{author}{\bibinfo {author} {\bibfnamefont{F.~C.}\ \bibnamefont{Alcaraz}}, \bibinfo
  {author} {\bibfnamefont{M.~N.}\ \bibnamefont{Barber}}, \bibinfo {author}
  {\bibfnamefont{M.~T.}\ \bibnamefont{Batchelor}}, \bibinfo {author}
  {\bibfnamefont{R.~J.}\ \bibnamefont{Baxter}},\ and\ \bibinfo {author}
  {\bibfnamefont{G.~R.~W.}\ \bibnamefont{Quispel}},\ }%
  \bibfield{journal}{\Doi{10.1088/0305-4470/20/18/038}{\bibinfo {journal} {J. Phys. A}}\ }%
  \textbf{\bibinfo {volume} {20}},\ \bibinfo {pages} {6397} (\bibinfo {year} {1987});
  \bibfield{author}{\bibinfo {author} {\bibfnamefont{V.}\ \bibnamefont{Murg}},\ \bibinfo
  {author} {\bibfnamefont{V.~E.}\ \bibnamefont{Korepin}}\ and\ \bibinfo
  {author} {\bibfnamefont{F.}\ \bibnamefont{Verstraete}},\ }%
  \bibinfo {note} {\Eprint{http://arxiv.org/abs/1201.5627}{arXiv:1201.5627}}
  \bibAnnoteFile{NoStop}{Alcaraz1987}%
\bibitem{Hirano2008}%
  \BibitemOpen
  \bibfield{author}{\bibinfo {author} {\bibfnamefont{T.}~\bibnamefont{Hirano}}, \bibinfo {author}
  {\bibfnamefont{H.}~\bibnamefont{Katsura}},\ and\ \bibinfo {author}
  {\bibfnamefont{Y.}~\bibnamefont{Hatsugai}},\ }%
  \bibfield{journal}{\Doi{10.1103/PhysRevB.77.094431}{\bibinfo {journal} {Phys. Rev. B}}\ }%
  \textbf{\bibinfo {volume} {77}},\ \bibinfo {pages} {094431} (\bibinfo {year}{2008})%
  \bibAnnoteFile{NoStop}{Hirano2008}%
\bibitem{Jiang2010}%
  \BibitemOpen
  \bibfield{author}{\bibinfo {author} {\bibfnamefont{H.-C.}\ \bibnamefont{Jiang}}, \bibinfo
  {author} {\bibfnamefont{S.}~\bibnamefont{Rachel}}, \bibinfo {author}
  {\bibfnamefont{Z.-Y.}\ \bibnamefont{Weng}}, \bibinfo {author}
  {\bibfnamefont{S.-C.}\ \bibnamefont{Zhang}},\ and\ \bibinfo {author}
  {\bibfnamefont{Z.}~\bibnamefont{Wang}},\ }%
  \bibfield{journal}{\Doi{10.1103/PhysRevB.82.220403}{\bibinfo {journal} {Phys. Rev. B}}\ }%
  \textbf{\bibinfo {volume} {82}},\ \bibinfo {pages} {220403}(R) (\bibinfo {year}{2010})%
  \bibAnnoteFile{NoStop}{Jiang2010}%
\bibitem{Zang2010}%
  \BibitemOpen
  \bibfield{author}{\bibinfo {author} {\bibfnamefont{J.}~\bibnamefont{Zang}}, \bibinfo {author}
  {\bibfnamefont{H.-C.}\ \bibnamefont{Jiang}}, \bibinfo {author}
  {\bibfnamefont{Z.-Y.}\ \bibnamefont{Weng}},\ and\ \bibinfo {author}
  {\bibfnamefont{S.-C.}\ \bibnamefont{Zhang}},\ }%
  \bibfield{journal}{\Doi{10.1103/PhysRevB.81.224430}{\bibinfo {journal} {Phys. Rev. B}}\ }%
  \textbf{\bibinfo {volume} {81}},\ \bibinfo {pages} {224430} (\bibinfo {year}{2010})%
  \bibAnnoteFile{NoStop}{Zang2010}%
\bibitem{Zheng2011}%
  \BibitemOpen
  \bibfield{author}{%
  \bibinfo {author} {\bibfnamefont{D.}~\bibnamefont{Zheng}}, \bibinfo {author}
  {\bibfnamefont{G.-M.}\ \bibnamefont{Zhang}}, \bibinfo {author}
  {\bibfnamefont{T.}~\bibnamefont{Xiang}},\ and\ \bibinfo {author}
  {\bibfnamefont{D.-H.}\ \bibnamefont{Lee}},\ }%
  \bibfield{journal}{\Doi{10.1103/PhysRevB.83.014409}{\bibinfo {journal} {Phys. Rev. B}}\ }%
  \textbf{\bibinfo {volume} {83}},\ \bibinfo {pages} {014409} (\bibinfo {year}{2011})%
  \bibAnnoteFile{NoStop}{Zheng2011}%
\bibitem{Tu2011}%
  \BibitemOpen
  \bibfield{author}{%
  \bibinfo {author} {\bibfnamefont{H.-H.}\ \bibnamefont{Tu}}\ and\ \bibinfo
  {author} {\bibfnamefont{R.}~\bibnamefont{Or\'us}},\ }%
  \bibfield{journal}{\Doi{10.1103/PhysRevB.84.140407}{\bibinfo {journal} {Phys. Rev. B}}\ }%
  \textbf{\bibinfo {volume} {84}},\ \bibinfo {pages} {140407}(R) (\bibinfo {year}{2011})%
  \bibAnnoteFile{NoStop}{Tu2011}%
\bibitem{Pollmann2012}%
  \BibitemOpen
  \bibfield{author}{\bibinfo {author} {\bibfnamefont{F.}~\bibnamefont{Pollmann}}, \bibinfo
  {author} {\bibfnamefont{E.}~\bibnamefont{Berg}}, \bibinfo {author}
  {\bibfnamefont{A.~M.}\ \bibnamefont{Turner}},\ and\ \bibinfo {author}
  {\bibfnamefont{M.}~\bibnamefont{Oshikawa}},\ }%
  \bibfield{journal}{\Doi{10.1103/PhysRevB.85.075125}{\bibinfo {journal} {Phys. Rev. B}}\ }%
  \textbf{\bibinfo {volume} {85}},\ \bibinfo {pages} {075125} (\bibinfo {year} {2012})%
  \bibAnnoteFile{NoStop}{Pollmann2012}%
\bibitem{Lou2000}%
  \BibitemOpen
  \bibfield{author}{\bibinfo {author} {\bibfnamefont{J.}~\bibnamefont{Lou}}, \bibinfo {author}
  {\bibfnamefont{S.}~\bibnamefont{Qin}},\ and\ \bibinfo {author}
  {\bibfnamefont{Z.}~\bibnamefont{Su}},\ }%
  \bibfield{journal}{\Doi{10.1103/PhysRevB.62.13832}{\bibinfo {journal} {Phys. Rev. B}}\ }%
  \textbf{\bibinfo {volume} {62}},\ \bibinfo {pages} {13832} (\bibinfo {year}{2000});
  \bibfield{author}{\bibinfo {author} {\bibfnamefont{F.}~\bibnamefont{Heidrich-Meisner}},
  \bibinfo {author} {\bibfnamefont{I.~A.}\ \bibnamefont{Sergienko}}, \bibinfo
  {author} {\bibfnamefont{A.~E.}\ \bibnamefont{Feiguin}},\ and\ \bibinfo
  {author} {\bibfnamefont{E.~R.}\ \bibnamefont{Dagotto}},\ }%
  \bibfield{journal}{\Doi{10.1103/PhysRevB.75.064413}{\bibinfo {journal} {\prb}}\ }%
  \textbf{\bibinfo {volume} {75}},\ \bibinfo {pages} {064413} (\bibinfo {year}{2007});
  \bibfield{author}{\bibinfo {author} {\bibfnamefont{Y.}~\bibnamefont{Zhao}},
  \bibinfo {author} {\bibfnamefont{S.-S.}\ \bibnamefont{Gong}}, \bibinfo
  {author} {\bibfnamefont{W.}\ \bibnamefont{Li}},\ and\ \bibinfo
  {author} {\bibfnamefont{G.}\ \bibnamefont{Su}},\ }%
  \bibfield{journal}{\Doi{10.1063/1.3413931}{\bibinfo {journal} {Appl. Phys. Lett.}}\ }%
  \textbf{\bibinfo {volume} {96}},\ \bibinfo {pages} {162503} (\bibinfo {year}{2010})%
  \bibAnnoteFile{NoStop}{Lou2000}%
\bibitem{PCChen2012}%
  \BibitemOpen
  \bibfield{author}{\bibinfo {author} {\bibfnamefont{P.}~\bibnamefont{Chen}}, \bibinfo {author}
  {\bibfnamefont{Z.-L.}\ \bibnamefont{Xue}}, \bibinfo {author}
  {\bibfnamefont{I.~P.}\ \bibnamefont{McCulloch}}, \bibinfo {author}
  {\bibfnamefont{M.-C.}\ \bibnamefont{Chung}},\ and\ \bibinfo {author}
  {\bibfnamefont{S.-K.}\ \bibnamefont{Yip}},\ }%
  \bibfield{journal}{\Doi{10.1103/PhysRevA.85.011601}{\bibinfo {journal} {Phys. Rev. A}}\ }%
  \textbf{\bibinfo {volume} {85}},\ \bibinfo {pages} {011601}(R) (\bibinfo {year}{2012});
  \bibfield{author}{\bibinfo {author} {\bibfnamefont{H.}~\bibnamefont{Nonne}}, \bibinfo {author}
  {\bibfnamefont{P.}\ \bibnamefont{Lecheminant}}, \bibinfo {author}
  {\bibfnamefont{S.}\ \bibnamefont{Capponi}}, \bibinfo {author}
  {\bibfnamefont{G.}\ \bibnamefont{Roux}},\ and\ \bibinfo {author}
  {\bibfnamefont{E.}\ \bibnamefont{Boulat}},\ }%
  \bibfield{journal}{\Doi{10.1103/PhysRevB.84.125123}{\bibinfo {journal} {\prb}}\ }%
  \textbf{\bibinfo {volume} {84}},\ \bibinfo {pages} {125123} (\bibinfo {year}{2011})%
  \bibAnnoteFile{NoStop}{PCChen2012}%
\bibitem{Haldane}%
  \BibitemOpen
  \bibfield{author}{\bibinfo {author} {\bibfnamefont{F.~D.~M.}\ \bibnamefont{Haldane}},\ }%
  \bibfield{journal}{\Doi{10.1016/0375-9601(83)90631-X}{\bibinfo {journal} {Phys. Lett. A}}\ }%
  \textbf{\bibinfo {volume} {93}},\ \bibinfo {pages} {464 } (\bibinfo {year}{1983});
  \BibitemOpen
  \bibfield{author}{\bibinfo {author} {\bibfnamefont{F.~D.~M.}\ \bibnamefont{Haldane}},\ }%
  \bibfield{journal}{\Doi{10.1103/PhysRevLett.50.1153}{\bibinfo {journal} {Phys. Rev. Lett.}}\ }%
  \textbf{\bibinfo {volume} {50}},\ \bibinfo {pages} {1153} (\bibinfo {year}{1983})%
  \bibAnnoteFile{NoStop}{Haldane}%
\bibitem{Qin2003}%
  \BibitemOpen
  \bibfield{author}{\bibinfo {author} {\bibfnamefont{S.}~\bibnamefont{Qin}}, \bibinfo {author}
  {\bibfnamefont{X.}~\bibnamefont{Wang}},\ and\ \bibinfo {author}
  {\bibfnamefont{L.}~\bibnamefont{Yu}},\ }%
  \bibfield{journal}{\Doi{10.1103/PhysRevB.56.R14251}{\bibinfo {journal} {\prb}}\ }%
  \textbf{\bibinfo {volume} {56}},\ \bibinfo {pages} {R14251} (\bibinfo {year}{1997});
  \bibfield{author}{\bibinfo {author} {\bibfnamefont{S.}~\bibnamefont{Qin}}, \bibinfo {author}
  {\bibfnamefont{Y.-L.}~\bibnamefont{Liu}},\ and\ \bibinfo {author}
  {\bibfnamefont{L.}~\bibnamefont{Yu}},\ }%
  \bibfield{journal}{\Doi{10.1103/PhysRevB.55.2721}{\bibinfo {journal} {Phys. Rev. B}}\ }%
  \textbf{\bibinfo {volume} {55}},\ \bibinfo {pages} {2721} (\bibinfo {year}{1997});
  \bibfield{author}{\bibinfo {author} {\bibfnamefont{X.}~\bibnamefont{Wang}}, \bibinfo {author}
  {\bibfnamefont{S.}~\bibnamefont{Qin}},\ and\ \bibinfo {author}
  {\bibfnamefont{L.}~\bibnamefont{Yu}},\ }%
  \bibfield{journal}{\Doi{10.1103/PhysRevB.60.14529}{\bibinfo {journal} {\prb}}\ }%
  \textbf{\bibinfo {volume} {60}},\ \bibinfo {pages} {14529} (\bibinfo {year}{1999});
  \bibfield{author}{\bibinfo {author} {\bibfnamefont{S.}~\bibnamefont{Qin}}, \bibinfo {author}
  {\bibfnamefont{J.}\ \bibnamefont{Lou}}, \bibinfo {author}
  {\bibfnamefont{L.}\ \bibnamefont{Sun}},\ and\ \bibinfo {author}
  {\bibfnamefont{C.}\ \bibnamefont{Chen}},\ }%
  \bibfield{journal}{\Doi{10.1103/PhysRevLett.90.067202}{\bibinfo {journal} {\prl}}\ }%
  \textbf{\bibinfo {volume} {90}},\ \bibinfo {pages} {067202} (\bibinfo {year}{2003})%
  \bibAnnoteFile{NoStop}{Qin2003}%
\bibitem{Schollwock1995}%
  \BibitemOpen
  \bibfield{author}{\bibinfo {author} {\bibfnamefont{U.}~\bibnamefont{Schollw\"ock}}\ and\
  \bibinfo {author} {\bibfnamefont{T.}~\bibnamefont{Jolic\oe{}ur}},\ }%
  \bibfield{journal}{\Doi{10.1209/0295-5075/30/8/009}{\bibinfo {journal} {Europhys. Lett.}}\ }%
  \textbf{\bibinfo {volume} {30}},\ \bibinfo {pages} {493} (\bibinfo {year} {1995})%
  \bibAnnoteFile{NoStop}{Schollwock1995}%
\bibitem{Schollwock1996}%
  \BibitemOpen
  \bibfield{author}{\bibinfo {author} {\bibfnamefont{U.}~\bibnamefont{Schollw\"ock}}, \bibinfo
  {author} {\bibfnamefont{O.}~\bibnamefont{Golinelli}},\ and\ \bibinfo {author}
  {\bibfnamefont{T.}~\bibnamefont{Jolic\oe{}ur}},\ }%
  \bibfield{journal}{\Doi{10.1103/PhysRevB.54.4038}{\bibinfo {journal} {Phys. Rev. B}}\ }%
  \textbf{\bibinfo {volume} {54}},\ \bibinfo {pages} {4038} (\bibinfo {year} {1996})%
  \bibAnnoteFile{NoStop}{Schollwock1996}%
\bibitem{noID}%
  \BibitemOpen
  \bibfield{author}{\bibinfo {author} {\bibfnamefont{H.}~\bibnamefont{Aschauer}}\ and\ \bibinfo
  {author} {\bibfnamefont{U.}~\bibnamefont{Schollw\"ock}},\ }%
  \bibfield{journal}{\Doi{10.1103/PhysRevB.58.359}{\bibinfo {journal} {Phys. Rev. B}}\ }%
  \textbf{\bibinfo {volume} {58}},\ \bibinfo {pages} {359} (\bibinfo {year} {1998})%
  \bibAnnoteFile{NoStop}{noID}%
\bibitem{Qin1995}%
  \BibitemOpen
  \bibfield{author}{\bibinfo {author} {\bibfnamefont{S.}~\bibnamefont{Qin}}, \bibinfo {author}
  {\bibfnamefont{S.}\ \bibnamefont{Liang}}, \bibinfo {author}
  {\bibfnamefont{Z.}\ \bibnamefont{Su}},\ and\ \bibinfo {author}
  {\bibfnamefont{L.}\ \bibnamefont{Yu}},\ }%
  \bibfield{journal}{\Doi{10.1103/PhysRevB.52.R5475}{\bibinfo {journal} {Phys. Rev. B}}\ }%
  \textbf{\bibinfo {volume} {52}},\ \bibinfo {pages} {R5475} (\bibinfo {year}{1995});
  \bibfield{author}{\bibinfo {author} {\bibfnamefont{M.}~\bibnamefont{Kumar}}, \bibinfo {author}
  {\bibfnamefont{S.}\ \bibnamefont{Ramasesha}},\ and\ \bibinfo {author}
  {\bibfnamefont{Z.~G.}\ \bibnamefont{Soos}},\ }%
  \bibfield{journal}{\Doi{10.1103/PhysRevB.79.035102}{\bibinfo {journal} {\prb}}\ }%
  \textbf{\bibinfo {volume} {79}},\ \bibinfo {pages} {035102} (\bibinfo {year}{2009})
  \bibAnnoteFile{NoStop}{Qin1995}%
\bibitem{note}
  \BibitemOpen
  \bibinfo {note} {The new parity DMRG proposed in this paper is not an improvement
                   for PBC algorithm. However, it provides an easy way to obtain accurate results
                   with parity quantum numbers for large sizes. The new parity DMRG may not be the
                   most efficient algorithm, but it is the only reliable parity algorithm beyond
                   Exact Diagonalization.}
  \bibAnnoteFile{Stop}{note}%
\bibitem{Verstraete2004}%
  \BibitemOpen
  \bibfield{author}{\bibinfo {author} {\bibfnamefont{F.}~\bibnamefont{Verstraete}}, \bibinfo {author}
  {\bibfnamefont{D.}~\bibnamefont{Porras}},\ and\ \bibinfo {author}
  {\bibfnamefont{J.~I.}~\bibnamefont{Cirac}},\ }%
  \bibfield{journal}{\Doi{10.1103/PhysRevLett.93.227205}{\bibinfo {journal} {\prl}}\ }%
  \textbf{\bibinfo {volume} {93}},\ \bibinfo {pages} {227205} (\bibinfo {year}{2004})
  \bibAnnoteFile{NoStop}{Verstraete2004}%
\bibitem{Pippan2010}%
  \BibitemOpen
  \bibfield{author}{\bibinfo {author} {\bibfnamefont{P.}~\bibnamefont{Pippan}}, \bibinfo {author}
  {\bibfnamefont{S.~R.}~\bibnamefont{White}},\ and\ \bibinfo {author}
  {\bibfnamefont{H.~G.}~\bibnamefont{Evertz}},\ }%
  \bibfield{journal}{\Doi{10.1103/PhysRevB.81.081103}{\bibinfo {journal} {\prb}}\ }%
  \textbf{\bibinfo {volume} {81}},\ \bibinfo {pages} {081103}(R) (\bibinfo {year}{2010})%
  \bibAnnoteFile{NoStop}{Pippan2010}%
\end{thebibliography}
\end{document}